\documentclass[aps,prb,floats,twocolumn,10pt,floatfix,superscriptaddress]{revtex4-1}
\usepackage{graphicx}
\usepackage{lipsum, babel}
\usepackage{amssymb}
\usepackage{amsmath}
\usepackage{mathtools}
\newcommand\numberthis{\addtocounter{equation}{1}\tag{\theequation}}
\usepackage{color}
\usepackage[dvipsnames]{xcolor}
\usepackage{slantsc}
\usepackage{float}
\graphicspath{{./FIG_FINALI/}}
\usepackage{bbold}
\usepackage[caption=false]{subfig}
\usepackage{array}
\usepackage{braket}
\usepackage{array}
\newcolumntype{P}[1]{>{\centering\arraybackslash}p{#1}}
\newcolumntype{M}[1]{>{\centering\arraybackslash}m{#1}}
\usepackage{scrextend}
\usepackage{hyperref}
\usepackage{bookmark}
\hypersetup{backref,pdfpagemode=FullScreen,colorlinks=true,allcolors=blue}
\usepackage{leftidx}

\begin{document}

\title{Efficient core-excited state orbital perspective on calculating X-ray absorption transitions in determinant framework}
\author{Subhayan Roychoudhury}
\email{roychos@tcd.ie}
\affiliation{The Molecular Foundry, Lawrence Berkeley National Laboratory, Berkeley CA 94720, USA}
\author{David Prendergast}
\email{dgprendergast@lbl.gov}
\affiliation{The Molecular Foundry, Lawrence Berkeley National Laboratory, Berkeley CA 94720, USA}

\begin{abstract}
X-ray absorption spectroscopy (XAS) is an explicit probe of the unoccupied electronic structure of materials and an invaluable tool for fingerprinting various electronic properties and phenomena. Computational methods capable of simulating and analysing such spectra are therefore in high demand for complementing the experimental results and for extracting valuable insights therefrom. In particular, a recently proposed first-principles approach titled Many-Body XAS (MBXAS), which approximates the final (initial) state as a Slater determinant constructed from Kohn-Sham (KS) orbitals optimized in absence (presence) of the relevant core-hole has shown promising prospects in evaluating the transition amplitudes. In this article, we show that the MBXAS approach can be rederived using a transition operator expressed entirely in the basis of core-excited state KS orbitals and that this reformulation offers substantial practical and conceptual advantages. In addition to circumventing previous issues of convergence with respect to the number of unoccupied ground-state orbitals, the aforementioned representation reduces the computational expense by rendering the calculation of such orbitals unnecessary altogether. The reformulated approach also provides a direct pathway for comparing the many-body approximation with the so-called single-particle treatment and indicates the relative importance in observed XAS intensity of the relaxation of the valence occupied subspace induced by the core excitation. Finally, using the core-excited state basis, we define auxiliary orbitals for x-ray absorption and demonstrate their utility in explaining the spectral intensity by contrasting them with single-particle approximations to the excited state.
\end{abstract}

\maketitle

\section{Introduction}
Several major aspects of modern technology are heavily reliant on our ability to investigate and engineer the electronic structure of functional materials. Therefore, a considerable portion of the advancements in electronics and computing~\cite{doi:10.1146/annurev-matsci-071312-121630, MESSMER1985285, doi:10.1098/rsta.2013.0270}, energy storage~\cite{doi:10.1126/sciadv.abm2422, https://doi.org/10.1002/eem2.12056}, renewable energy harvesting~\cite{PhysRevLett.112.196401, doi:https://doi.org/10.1002/9783527813636.ch19}, drug discovery~\cite{Schaduangrat2020, Arodola2017,Ye2022}, discovery of magnetic materials~\cite{doi:10.1126/sciadv.1602241}, etc. can be attributed to the advancements in the theoretical and experimental research on electronic structure theory. On the experimental front, electronic structure of materials is usually explored by creating electronic excitations. Among such methods, X-ray absorption spectroscopy (XAS)~\cite{doi:10.1021/cr9900681,doi:10.1021/acs.chemrev.7b00213,PhysRevLett.96.215502,Li_2016,FREIWALD2004413,liu_weng_2016,C4CP05316H,C6CP02412B}, which offers an element-specific probe, typically with good resolution, has emerged as one of the most popular and effective techniques over the last few decades. Besides providing a relatively straightforward avenue for investigating, explicitly, the electronic structure (mostly of the unoccupied subspace) of materials, the XAS process is widely employed for fingerprinting various chemical attributes and phenomena of interest~\cite{https://doi.org/10.1002/admi.201300115,https://doi.org/10.1002/eem2.12119,HORIKAWA201233,doi:10.1021/acs.jpcc.5b07479,BRANCI2000363}. Within a quasielectron framework, the XAS process amounts to excitation of the core electrons of the desired atomic species to the unoccupied levels and measures, albeit typically through indirect means, the cross-section of such transitions. As with most experimental techniques, the interpretation of XAS data and the extraction of information related to electronic structure and properties therefrom is often facilitated by complementary theoretical and computational research~\cite{doi:10.1021/acsami.1c11970}. Development of efficient and accurate theoretical models for simulating XAS is, therefore, an active and dynamic field of research.

Among the various first-principles methods employed in computational electronic structure theory, Kohn-Sham (KS) density-functional theory (DFT) has witnessed unprecedented popularity in a myriad of applications owing to its favorable system-size scaling, modest computational requirements and typically reasonable accuracy. Consequently, in the field of X-ray spectroscopy, substantial effort has been invested in developing and applying theoretical models built upon KS DFT. For instance, the response based approaches, such as linear-response (LR) time-dependent DFT (TDDFT)~\cite{C002207A,doi:10.1021/jp050755y,doi:10.1063/5.0092987,doi:10.1021/ct3005613} or the Bethe-Salpeter equation (BSE)~\cite{PhysRevLett.80.794,PhysRevB.82.205104,PhysRevB.79.041102}, aim to evaluate the absorption cross-section from the response of the electron density to a change in the potential. Such response, in turn, is obtained most commonly by approximating the excited states as a linear combination of singles excitations of the KS ground state (GS). However, the response-based technique can prove to be expensive, especially when it involves explicit solution of the relevant equation (LR-TDDFT/BSE) within an all-electron calculation~\cite{PhysRevB.82.205104,PhysRevB.79.041102}. Additionally, the response-based approach, within the commonly used adiabatic approximation can incur substantial errors in core-level spectroscopy, due to the inadequate description of the electron-hole (e-h) interaction~\cite{PhysRevB.106.075133,doi:10.1063/5.0092987}. A viable inexpensive alternative is the constrained-occupation approach, whereby the excited state is modeled by maintaining a hole in the core system with the help of additional constraints. Interestingly, such a treatment can be used in a pseudopotential-based calculation by representing the core of the excited atom by a modified pseudopotential that encodes the effects of the core-hole~\cite{PhysRevB.66.195107,PhysRevLett.96.215502,PhysRevLett.118.096402}. The approximate KS orbitals corresponding to the core-excited state can then be obtained from a self-consistent field (SCF) calculation employing the aforementioned core-hole pseudopotential either within the Full Core-Hole (FCH) treatment, which minimizes the energy of the core-ionized system with a net positive charge or within the eXcited-electron and Core-Hole (XCH) treatment, which minimizes the energy of the neutral core-excited state.

Within the so-called single-particle approximation, one can estimate the absorption amplitude as that of a transition of a non-interacting electron from the core-orbital to the relevant unoccupied KS orbital. Such a simplified calculation, however, neglects the explicit effects of the relaxation of the occupied orbitals in response to the core-excitation, resulting often in severe inaccuracies in the absorption intensities.  To mitigate these errors, the recently proposed Many-Body X-ray Absorption Spectroscopy (MBXAS) method relies on two separate SCF calculations:
\begin{enumerate}
    \item a GS calculation generating the orbitals for the initial state, and
    \item a FCH/XCH calculation generating the orbitals for the core-excited state.
\end{enumerate}
Each state is then expressed as a Slater determinant (SD) composed of the corresponding orbitals and the transition amplitude is obtained, typically within the dipole approximation, as a many-body matrix element of the transition operator between the initial and the final state SDs. This determinant-based technique, which only requires simple pseudopotential-based KS DFT calculations and consequently retains their low computational cost and scaling, has been shown to significantly improve the peak-intensities of the single-particle approximation, especially for the O \textit{K}-edge of transition metal oxides, and to obtain very good agreement with the experimental spectra.

Building upon this development, in this article we reformulate the MBXAS formalism by recasting the relevant matrix elements with respect to the Full core-hole Basis (FB) which improves the computational efficiency, eliminates issues of convergence and enables direct comparison with the single-particle approximation. Additionally, we develop the concept of auxiliary orbitals for XAS which helps with the association of physically-motivated single-particle orbitals to the excited many-electron system.

The rest of the article is organized as follows. We introduce the reformulated treatment of MBXAS in the next section, which is followed by a prescription, inspired from Ref.~\cite{PhysRevB.97.205127}, for achieving computational simplification. The following subsection, where we present and discuss the results of our calculation on some representative systems, starts with a discussion, along with some illustrative examples, on the computational advantages of recasting the expressions to the core-hole basis set. This is then followed by a separation of the terms contributing to the MBXAS amplitude into categories of different physical origins which facilitate a direct comparison with the single-particle amplitude, as discussed in the next subsection. Finally, we move on to the discussion of auxiliary orbitals corresponding to X-ray excitation.

\section{Method}
\subsection{General Background}
Within a quasiparticle-inspired framework, X-ray absorption results in the neutral excitation of an electronic system typically from its ground state to an excited final state containing a core-hole and a conduction electron.
In order to simplify the terminology, without loss of generality, in the following, the term ``core'' is reserved only to denote the electron/hole corresponding to the specific orbital emptied as a consequence of X-ray absorption. Denoting by $a^\dagger_i$ ($a^\dagger_0$) the creation operator corresponding to the $i$-th valence level $\ket{\phi_i}$ (core level $\ket{\phi_0}$), the ground state (GS) of the system can be expressed as the Slater determinant (SD)
\begin{align}\label{GS}
    \ket{\Psi_{\rm{GS}}} = \left(\prod_{i=1}^N a^\dagger_i\right) a_0^\dagger\ket{\emptyset},
\end{align}
where $\ket{\emptyset}$ denotes the null state. In contrast, a final state can be constructed by populating the KS orbitals obtained from a SCF calculation performed in absence of the relevant core electron, which, in accordance with the standard convention, will be called a full core-hole (FCH) SCF hereafter. Using a $\sim$ to indicate quantities corresponding to such a core-excited SCF, the core-ionized system in its lowest energy state will be given by 
\begin{align}\label{FCH_STATE}
    \ket{\Psi_{\rm{FCH}}}=\left(\prod_{\mu=1}^N \tilde{a}^\dagger_\mu\right)\ket{\emptyset} ,
\end{align}
with the explicit absence of the core orbital, indexed as 0. Then, within the singles approximation, any neutral final state $\ket{\Psi_f}$ in which, in addition to the lowest $N$ valence orbitals, the conduction orbital $\ket{\tilde{\phi}_f} = \tilde{a}_f^\dagger\ket{\emptyset}$ is now occupied, can be constructed from $\ket{\Psi_{\rm{FCH}}}$ as
\begin{align}\label{FinalState}
 \ket{\Psi_f}= \ket{\Psi_{\rm{FCH}}^{+f}}=\tilde{a}_f^\dagger \ket{\Psi_{\rm{FCH}}}.
\end{align}

\subsection{MBXAS in core-excited state basis}

The process of electronic transition is governed by the many-body transition operator $\hat{O}$ which, in the earlier works, has been represented in terms of the GS orbitals as
\begin{align}
    \hat{O}=\sum_{i,j}\braket{\phi_i|\hat{o}|\phi_j}a^\dagger_i a_j.
\end{align}
with $\hat{o}$ standing for the single-particle transition operator and the sum extending over all pairs of orbitals of the GS system.

However, as with any operator, $\hat{O}$ can, in fact, be expanded in any complete single-particle Hilbert space representation. This prompts us to express the transition operator with respect to the FCH state orbitals as 
\begin{align*}\label{TransitionOperator}
    \hat{O} &{} = \sum_{l,h} \braket{\tilde{\phi}_l|\hat{o}|\tilde{\phi}_h} \tilde{a}_l^\dagger \tilde{a}_h\\
    &{} = \sum_{l,h} \tilde{o}_{lh} \tilde{a}_l^\dagger \tilde{a}_h,\numberthis
\end{align*}
which, as discussed in the following sections, offers substantial advantage in terms of computation and interpretation of the X-ray absorption spectra. By combining Eqs.~\ref{GS}, \ref{FCH_STATE}, \ref{FinalState}, and~\ref{TransitionOperator}, we obtain the expression for the transition amplitude
\begin{align*}\label{TransitionAmplitude}
A_{\rm{GS}\rightarrow f} &{} = \braket{\Psi_f|\hat{O}|\Psi_{\rm{GS}}}\\
     &{} = \sum_{l,h} \tilde{o}_{lh} \braket{0| \left(\prod_{\mu=1}^N \tilde{a}_\mu\right) \tilde{a}_f \tilde{a}_l^\dagger \tilde{a}_h \left(\prod_{i=1}^N a_i^\dagger \right) a_0^\dagger |0}\\
     &{} =  \sum_{l=1..N,f} \tilde{o}_{l} \braket{0| \left(\prod_{\mu=1}^N \tilde{a}_\mu\right) \tilde{a}_f \tilde{a}_l^\dagger \tilde{a}_0 \left(\prod_{i=1}^N a_i^\dagger \right)  a_0^\dagger|0}\\
     &{} =  \sum_{l=1..N,f} \tilde{o}_l \braket{\Psi_f|\tilde{a}_l^\dagger\tilde{a}_0|\Psi_{\rm{GS}}}, \numberthis
\end{align*}
where, for the sake of simplicity, we have used the notation $\tilde{o}_{l}=\tilde{o}_{l0}$. The third line follows from the observation that, in the sum over $h$, all terms except for that corresponding to $h=0$ (i.e, where $h$ is essentially the core-level) vanish. The approximation $\ket{\phi_0}\approx \ket{\tilde{\phi}_0}$ is justified due to the extremely localized nature of the core orbital~\cite{PhysRevB.106.075133}.

Now, for any $f>N$ and $l$ in the range $1..N$, we introduce the many-electron state 
\begin{align}\label{NonScFinalState}
    \ket{\Psi_{\rm{FCH}}^{+f-l+0}} = (-1)^{\gamma_l} \left(\tilde{a}_0^\dagger \tilde{a}_l \ket{\Psi_f}\right),
\end{align}
which has an index-ordered set of occupied orbitals, where the factor $(-1)^{\gamma_l}$, with $\gamma_l=(2N-l+1)$, originates from the anticommutation relations of the creation and annihilation operators. More specifically, in any expression involving the creation and annihilation operators, an exchange of two consecutive operators generates a factor of $-1$. Thus, given that, in order to obtain $\ket{\Psi_{\rm{FCH}}^{+f-l+0}}$ from $\ket{\Psi_f}$, the operators $\tilde{a}_l$ and $\tilde{a}_0^\dagger$ must be moved $N+1-l$ and $N$ places, respectively, a resultant factor of $(-1)^{2N+l-1}$ is introduced. Eq.~\ref{TransitionAmplitude} can then be rewritten as
\begin{equation}
\begin{aligned}\label{MBXAS_amplitude}
    A_{\rm{GS}\rightarrow f} = & \sum_{l=1..N} (-1)^{\gamma_l} \tilde{o}_l \braket{\Psi_{\rm{FCH}}^{+f-l+0}|\Psi_{\rm{GS}}} \\
    & + (-1)^{\gamma_f} \tilde{o}_f \braket{\Psi_{\rm{FCH}}^{+0}|\Psi_{\rm{GS}}},
\end{aligned}
\end{equation}
where $\gamma_f=N$, and we have used the notation $\ket{\Psi_{\rm{FCH}}^{+f-f+0}} = \ket{\Psi_{\rm{FCH}}^{+0}} = (-1)^N \tilde{a}_0^\dagger \tilde{a}_f\ket{\Psi_f}$. Now, noting that an overlap of two SDs can be expressed as a determinant composed of the constituent orbital overlaps, the term $\braket{\Psi_{\rm{FCH}}^{+f-l+0}|\Psi_{\rm{GS}}}$ 
can be expressed as
\begin{align}\label{OverlapSD}
    \braket{\Psi_{\rm{FCH}}^{+f-l+0}|\Psi_{\rm{GS}}} = \det \begin{bmatrix} \xi^*_{1,1} & \hdots & \xi^*_{1,N}\\ \vdots & \ddots & \vdots\\ \xi^*_{l-1,1} & \hdots & \xi^*_{l-1,N}\\ \xi^*_{l+1,1} & \hdots & \xi^*_{l+1,N}\\ \vdots & \ddots & \vdots\\ \xi^*_{N,1} & \hdots & \xi^*_{N,N}\\ \xi^*_{f,1} & \hdots & \xi^*_{f,N} \end{bmatrix},
\end{align}
where $\xi_{i,j}=\braket{\phi_j|\tilde{\phi}_i}$ denotes the overlap between the GS and the FCH KS orbitals. Thus, with the help of Eq.~\ref{MBXAS_amplitude} and~\ref{OverlapSD}, the MBXAS method approximates the transition amplitude for any final state $\ket{\Psi_{\rm{FCH}}^{+f}}$ from the results of two converged SCF calculations of KS DFT : one for the ground-state  and one for the core-ionized state.
\subsection{Computational simplification}

Introducing the notation
\begin{align}
    \alpha_q = \left[\xi_{q,1}\textrm{ } \xi_{q,2} \hdots \xi_{q,N}\right],
\end{align}
a vector of overlaps, we have
\begin{align}
    \braket{\Psi_{\rm{GS}}|\Psi_{\rm{FCH}}^{+f-l+0}} = \alpha_1 \wedge \hdots \wedge \alpha_{l-1}\wedge \alpha_{l+1} \wedge \hdots \wedge \alpha_N \wedge \alpha_f
\end{align}
Givn that this $N$-dimensional determinant defines a finite $N$-dimensional vector space, then, for any $f>N$, $\alpha_f$ must be linearly dependent on the original $N$ vectors, $\alpha_1 \dots \alpha_N$, which can be expressed in the form of a linear equation
\begin{align}\label{LinearEqForAlpha}
    \alpha_f = \sum_{k=1}^N \kappa_{f,k}\alpha_k,
\end{align}
with $\{\kappa_{f,k}\}$ denoting a set of unknown coefficients which can be computed by expressing the linear relations of Eq.~\ref{LinearEqForAlpha}, for $f$ in the range $N+1 \hdots M$, in the form of a matrix equation as
\begin{align}\label{MatrixEquation}
    \begin{bmatrix}
    \alpha_{N+1}\\ \vdots \\ \alpha_M
    \end{bmatrix} = \begin{bmatrix}
    \kappa_{N+1,1} & \hdots & \kappa_{N+1,N}\\
    \vdots & \ddots & \vdots \\
    \kappa_{M,1} & \hdots & \kappa_{M,N}
    \end{bmatrix} \begin{bmatrix}
    \alpha_1 \\ \vdots \\ \alpha_N
    \end{bmatrix}
\end{align}
Denoting the first, second and third matrices in Eq.~\ref{MatrixEquation} as $\rm{A'_{mat}}$, $\rm{K_{mat}}$, and $\rm{A_{mat}}$, respectively, $\rm{K_{mat}}$ can be calculated from the equation
\begin{align}
    \rm{K_{mat}} = \rm{A'_{mat}} . \left(\rm{A_{mat}}\right)^{-1}
\end{align}

Now, using the identity $a_i \wedge a_j = (- a_j \wedge a_i).\delta_{i,j}$, the complex-conjugate of the determinantal overlap of Eq.~\ref{OverlapSD} can be rewritten as
\begin{align*}
    &{}\braket{\Psi_{\rm{GS}}|\Psi_{\rm{FCH}}^{+f-l+0}}\\ &{}= \alpha_1 \wedge \hdots \wedge \alpha_{l-1}\wedge \alpha_{l+1} \wedge \hdots \wedge \alpha_N \wedge \left(\sum_{k=1}^N \kappa_{f,k}\alpha_k\right)\\
    &{} = (-1)^{N-l}\kappa_{f,l}\left(\alpha_1 \wedge \hdots \wedge \alpha_N\right)\\
    &{} = (-1)^{N-l}\kappa_{f,l} \rm{A_{det}}\numberthis ,
\end{align*}
where 
\begin{align}
    \rm{A_{det}} = \left(\alpha_1 \wedge \hdots \wedge \alpha_N\right) = \braket{\Psi_{\rm{GS}}|\Psi_{\rm{FCH}}^{+0}}.
    \label{ValenceOverlap}
\end{align}
Then, from Eq.~\ref{MBXAS_amplitude}, the transition amplitude can be calculated as
\begin{align}
    \braket{\Psi_f|\hat{O}|\Psi_{\rm{GS}}} &{} = (-1)^N \left[\tilde{o}_{f} - \sum_{l=1..N} \kappa^*_{f,l} \tilde{o}_{l} \right] \rm{A^*_{det}} .
\end{align}
\section{Discussions}

\subsection{Computational Advantage}
\begin{figure}
\centering
\includegraphics[width=0.42\textwidth]{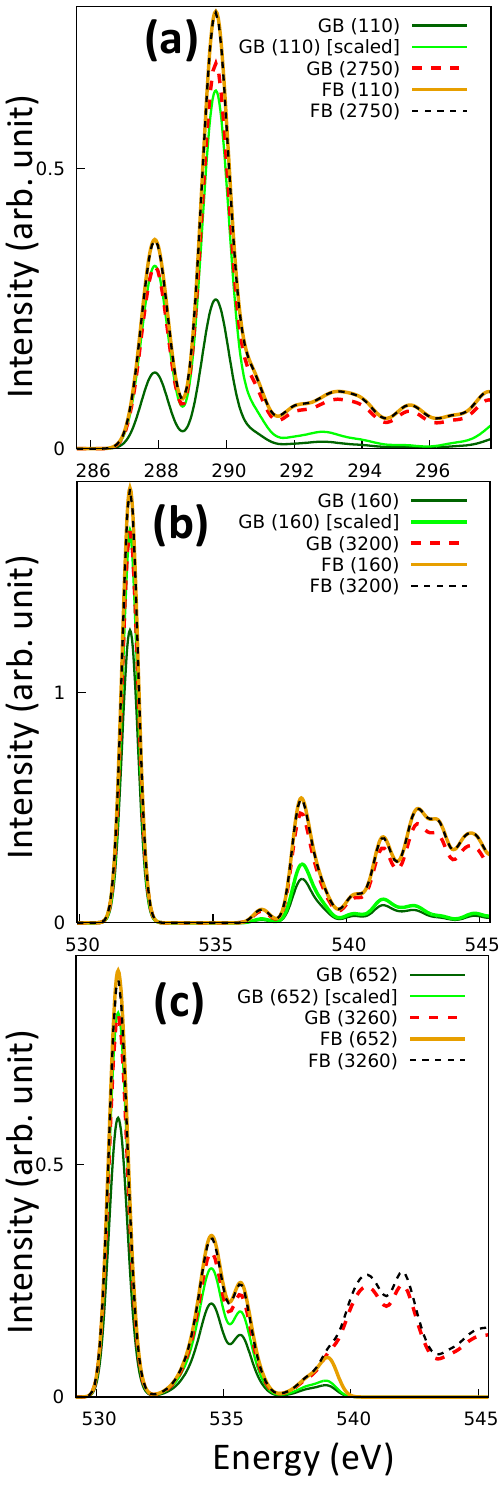}
\caption{(a) Methanol C \textit{K}-edge, (b) Acetone O \textit{K}-edge and (c) CuO O \textit{K}-edge absorption spectra. The number in the bracket denotes the total number of FCH KS orbitals (excluding the core orbitals) used in the calculation. The ``scaled'' curve in each panel is obtained by multiplying the dark-green one by a constant factor in order to match the first-peak intensity with that of the broken red one.}\label{GBA_FBA_ConvComp}
\end{figure}

It is instructive to compare the MBXAS formalism in the FB Approach as presented above with the original Ground state Basis (GB) Approach proposed in Ref.~\cite{PhysRevB.97.205127}. Both of the approaches aim to compute the transition amplitude from the identical expression $A_{\rm{GS}\rightarrow f} = \braket{\Psi_f|\hat{O}|\Psi_{\rm{GS}}}$ with the state $\ket{\Psi_{\rm{GS}}}$ and $\ket{\Psi_{f}}$ being given by Eq.~\ref{GS} and Eq.~\ref{FinalState}, respectively. Therefore, at convergence, the spectra calculated using the FB and the GB approaches should, in principle, be the same. In practice, however, there can be substantial differences. 

In the GB treatment, the transition amplitude within the singles approximation is simplified to
\begin{align*}\label{GBA_AMPLITUDE}
    A_{\rm{GS}\rightarrow f} &{}= \sum_{c>N} \braket{\Psi_{\rm{FCH}}^{+f}|\Psi_{\rm{GS}}^{+c-0}} \braket{\Psi_{\rm{GS}}^{+c-0}|\hat{O}|\Psi_{\rm{GS}}} \\&{}= \sum_{c>N} o_c \braket{\Psi_{\rm{FCH}}^{+f}|\Psi_{\rm{GS}}^{+c-0}},\numberthis
\end{align*}
where $o_c=\braket{\phi_c|\hat{o}|\phi_0}$ and 
\begin{align}
  \ket{\Psi_{\rm{GS}}^{+c-0}}=(-1)^Na_c^\dagger a_0  \ket{\Psi_{\rm{GS}}},  
\end{align}
The relevant many-body overlap is then given by
\begin{align}
    \braket{\Psi_{\rm{FCH}}^{+f}|\Psi_{\rm{GS}}^{+c-0}} = (-1)^N \begin{bmatrix} \xi^*_{1,1} & \hdots & \xi^*_{1,N} & \xi^*_{1,c}\\
    \vdots & \ddots & \vdots & \vdots\\ \xi^*_{N,1} & \hdots & \xi^*_{N,N} & \xi^*_{N,c}\\ \xi^*_{f,1} & \hdots & \xi^*_{f,N} & \xi^*_{f,c} \end{bmatrix},
\end{align}

Note that, in principle, the sum in Eq.~\ref{GBA_AMPLITUDE} extends over all unoccupied GS orbitals - bound and unbound both.
Thus, if one fails to include any GS unoccupied orbital which has a non-zero overlap with any of the valence occupied FCH-state orbitals (i.e. $\ket{\tilde{\phi}_1},\hdots,\ket{\tilde{\phi}_N}$), then, in principle, $A_{\rm{GS}\rightarrow f}$ will be under-converged for any $f>N$. Additionally, if, for a given $f>N$, $\ket{\tilde{\phi}_f}$ overlaps with any missing unoccupied GS orbitals, then the corresponding $A_{\rm{GS}\rightarrow f}$ will be under-converged.
Therefore, convergence in Eq.~\ref{GBA_AMPLITUDE} can be particularly difficult to achieve if one or more of these FCH-state orbitals has appreciable overlap with the unbound unoccupied orbitals of the GS~\cite{doi:10.1139/cjp-2014-0726}.

As an aside, we may note that similar convergence issues should plague any computational method for simulating XAS that is rooted in a GS orbital representation of the XAS transition amplitude. Prominent examples include both the aforementioned LR-TDDFT~\cite{C002207A,doi:10.1021/jp050755y,doi:10.1063/5.0092987,doi:10.1021/ct3005613} and BSE~\cite{PhysRevLett.80.794,PhysRevB.82.205104,PhysRevB.79.041102} approaches, which are both expressed within a GS orbital basis that includes unoccupied orbitals.

As illustrative examples, in Fig.~\ref{GBA_FBA_ConvComp}, we present comparative studies of the GB and the FB treatments for the C \textit{K}-edge spectrum of methanol ($\rm{CH_3OH}$), O \textit{K}-edge spectrum of acetone ($\rm{C_3H_6O}$) and the O \textit{K}-edge spectrum of cupric oxide ($\rm{CuO}$).
The GB spectra are seen to be heavily dependent on the number of unoccupied GS orbitals employed in the calculation. The fact that the large GB spectra do not coincide with the scaled small GB counterparts indicates that the under-converged GB spectra differ inherently in their lineshapes.

Since the sum in Eq.~\ref{TransitionAmplitude} is limited to a finite number $(N+1)$ of terms anyway, this complication is circumvented entirely in the FB treatment as evident from Fig.~\ref{GBA_FBA_ConvComp}, which shows, as expected, that the FB spectra do not depend on the number of orbitals included in the calculation, as long as all FCH-state orbitals within the desired energy range are included. In addition to not suffering from convergence issues, for any given energy range of the spectrum, the FB treatment is computationally favourable over the GB treatment since, unlike the latter, the former does not require the computation of any unoccupied GS orbital. 

For a simulation involving $N_{\rm{tot}}$ number of GS orbitals (including the valence occupied orbitals as well as the unoccupied ones), within the GB approach, a practical MBXAS calculation requires the execution of the following steps separately for each of the GS and the FCH systems:
\begin{enumerate}
    \item a SCF calculation to obtain the electron density,
    \item a non self-consistent field (NSCF) calculation to obtain the unoccupied KS orbitals, which
    \item for computational simplification, are often transformed into an optimal basis-representation to facilitate faster Brillouin zone integration.
\end{enumerate}
Additionally, for the GS system, we
\begin{enumerate}
\setcounter{enumi}{3}
    \item calculate the single-particle transition matrix elements $o_c$ for all unoccupied GS orbitals $\ket{\phi_c}$.
\end{enumerate}
This is followed by
\begin{enumerate}
\setcounter{enumi}{4}
    \item calculation of the $(N_{\rm{tot}} \times N_{\rm{tot}})$ overlap matrix between the ground and the FCH state orbitals,
\end{enumerate}
and finally
\begin{enumerate}
\setcounter{enumi}{5}
    \item the computation of the MBXAS amplitudes from Eq.~\ref{GBA_AMPLITUDE}.
\end{enumerate}

On the other hand, in the FB treatment, for the GS system, step (2) is not required at all while step (3) can become substantially cheaper due to the absence of the unoccupied orbitals. The same is true for step (5) which now calculates the $(N \times N_{\rm{tot}})$ overlap matrix $\xi_{i,j}$ $\forall i\leq N_{\rm{tot}}\textrm{ and }j\leq N$ and for step (6) which, now follows Eq.~\ref{MBXAS_amplitude} and thereby involves a sum over $(N+1)$ terms. Note that, in the FB approach, step (4) is replaced by a calculation of the single-particle transition matrix elements $\tilde{o}_l$ over \textit{all} of the FCH-state orbitals, occupied and unoccupied alike, and therefore, should be a little more expensive than the corresponding GB calculation. However, typically, for a calculation involving a large number of unoccupied orbitals, steps (2) and (3) are, by a large margin, the most expensive ones computationally. Therefore, assuming that, for a given system, they have comparable costs for the GS and the FCH state, the FB approach is expected to reduce the total computational cost approximately by half. This, in fact, can be seen from Table~\ref{TableI}, which shows the estimates of the computational time required for methanol C \textit{K}-edge, acetone O \textit{K}-edge and CuO O \textit{K}-edge calculations with a relatively large number of unoccupied orbitals.

\begin{table}[h]
\centering
\begin{tabular}{|M{3.8cm}|M{1.5cm}|M{1.5cm}|}
    \hline 
    System & Core hours (GB) & Core hours (FB)\\
    \hline
    Methanol C \textit{K} (2750) & 95 & 46\\
    
    Acetone O \textit{K} (3200) & 166 & 82\\
    
    CuO O \textit{K} (3260) & 1056 & 597\\
    
    \hline \hline
    \end{tabular}
\caption{Table showing for methanol C \textit{K}-edge, acetone O \textit{K}-edge and CuO O \textit{K}-edge, the estimates of core hours used in the XAS calculations using the GB and the FB approaches. The number in the bracket denotes the total number of FCH KS orbitals (excluding the core orbitals) used in the calculation.}\label{TableI}
    \end{table}

\subsection{Separation of Spectral Contributions}
Note from Eq.~\ref{MBXAS_amplitude}, that the MBXAS amplitude $A_{\rm{GS} \rightarrow f}$ is essentially a linear-combination, for $l=1, \hdots, N$ and $l=f$, of single-particle transition amplitudes $\tilde{o}_l$ from the core orbital to the orbital $\ket{\tilde{\phi}_l}$, with the factor $\braket{\Psi_{\rm{FCH}}^{+f-l+0}|\Psi_{\rm{GS}}}$ serving as the coefficient.
We can naturally separate these terms according to their physical origins as follows:
\begin{enumerate}
    \item The $l=f$ term essentially corresponds to the expected quasiparticle transition, from the core orbital to orbital $\ket{\tilde{\phi}_f}$, which is formally unoccupied in $\ket{\Psi_{\rm{FCH}}}$, but indirectly includes some perturbative effects of the presence of the core-hole (since it is calculated based on the core-excited SCF). 
    \item The $l=1, \hdots, N$ terms accounts for the collective response of the rest of the electron density to the perturbation caused by the core hole by considering single-particle contributions from each of the occupied valence orbitals of $\ket{\Psi_{\rm{FCH}}}$.
    
\end{enumerate}

This separation is just a natural consequence of the many-body transition operator coupling the $N$-electron ground and core-excited states. However, another way to understand it is to consider the equivalence of absorption amplitude to the (stimulated) emission amplitude. We can rewrite Eq.~\ref{MBXAS_amplitude} as
\begin{align*}
    &{} A_{\rm{GS}\rightarrow f}\\ &{}= (-1)^{\gamma_l} \left[\sum_{l=1,\hdots,N,f}\braket{\Psi_{\rm{FCH}}^{+f}|\hat{O}|\Psi_{\rm{FCH}}^{+f-l+0}}\braket{\Psi_{\rm{FCH}}^{+f-l+0}|\Psi_{\rm{GS}}}\right].\numberthis
\end{align*}
Now we can interpret this as a de-excitation from $\ket{\Psi_{\rm{FCH}}^{+f}}$ to $\ket{\Psi_{\rm{GS}}}$, where the $l$-th representative term can be recognized as the amplitude $\braket{\Psi_{\rm{FCH}}^{+f}|\hat{O}|\Psi_{\rm{FCH}}^{+f-l+0}}$ of a \textit{non-selfconsistent} de-excitation from the state $\ket{\Psi_{\rm{FCH}}^{+f}}$ to the core-filled state $\ket{\Psi_{\rm{FCH}}^{+f-l+0}}$, multiplied, for the sake of restoring self-consistency, by the projection of this non-self consistent state onto the actual final state $\ket{\Psi_{\rm{GS}}}$. Thus, each of the $(N+1)$ terms in the sum corresponds to de-excitation of an electron from one of the occupied levels of $\ket{\Psi_{\rm{FCH}}^{+f}}$ to the core level.

\subsection{Comparison with the single-particle formalism}
Armed with the numerically converged FB treatment, we can now examine the relative importance of its various contributions. According to the expansion shown in Eq.~\ref{MBXAS_amplitude}, $\tilde{o}_l$ is one among the $N+1$ terms ($l=1, \hdots, N,f$) contributing to the transition amplitude. This affords a straightforward avenue for comparison of the MBXAS formalism with the more common single-particle formalism of XAS, in which the transition amplitude $A_{\rm{GS}\rightarrow f}$ is approximated simply by $\tilde{o}_f$. 

Mathematically, it can be shown that this truly single-particle approximation would be true if the valence occupied subspace of the ground and the FCH state were identical. Then $\braket{\Psi_{\rm{FCH}}^{+f-l+0}|\Psi_{\rm{GS}}} = \delta_{f,l}$ and consequently, $A_{\rm{GS}\rightarrow f}=\tilde{o}_f$ (possibly with an inconsequential change in sign). We could now ask whether the degree to which the valence occupied subspaces differ is a useful metric to indicate a larger difference between $A_{\rm{GS}\rightarrow f}$ and $\tilde{o}_f$. We explore this possibility through a detailed analysis of the first peak of the C and O \textit{K}-edge absorption spectra of an isolated methanol molecule.

\begin{figure}
\centering
\includegraphics[width=0.49\textwidth]{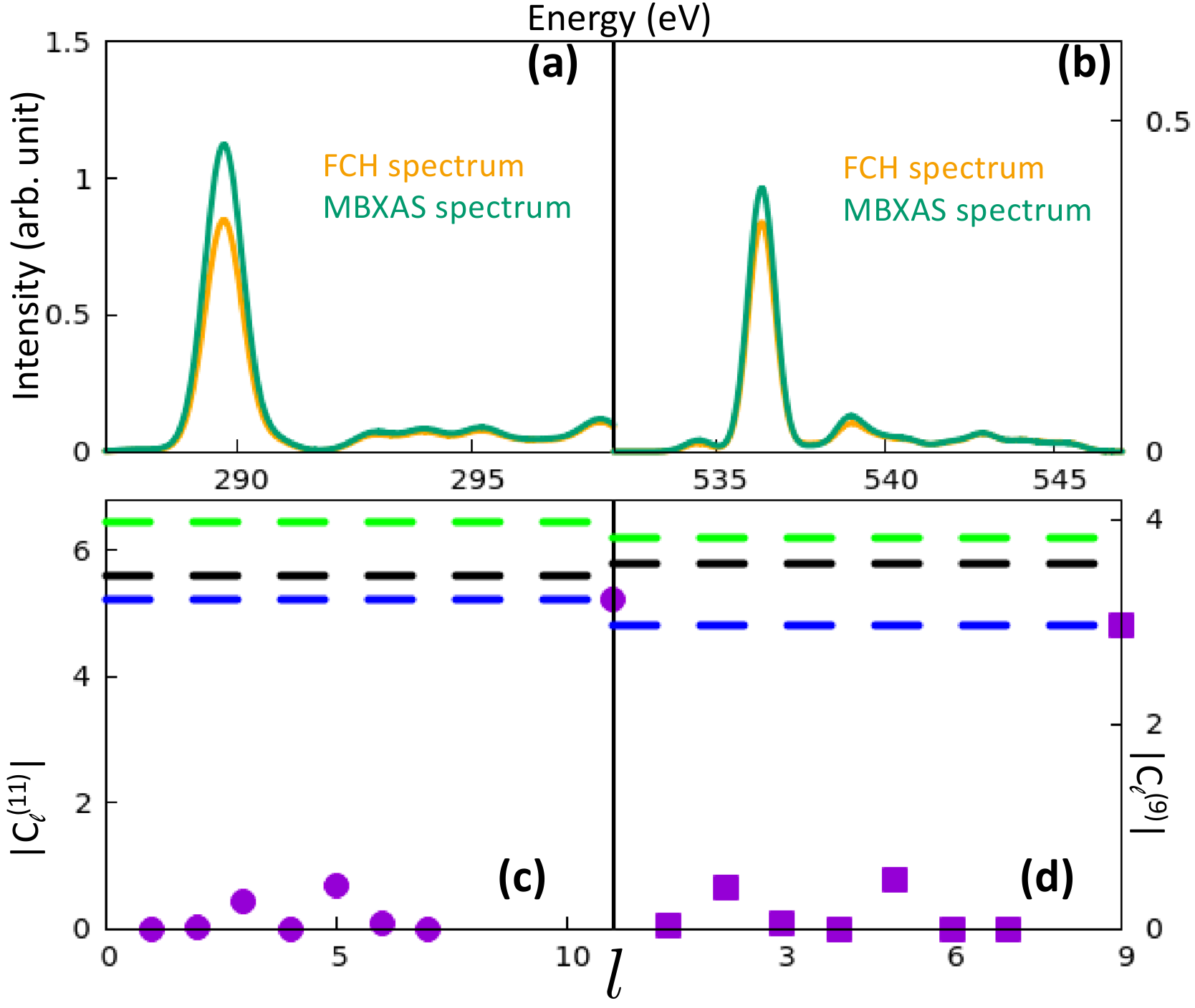}
\caption {Panels (a) and (b) show, for the methanol molecule, the C \textit{K}-edge XAS spectrum for incident polarization along the x-axis and the O \textit{K}-edge XAS spectrum for incident polarization along the y-axis, respectively (see molecular polarizaton in Fig.~\ref{MethanolOrbitalPlot}). Panels (c) and (d) analyse, in detail, the orbital contributions responsible for MBXAS transition amplitude of the most intense peaks seen in panels (a) and (b), respectively. In panel (c) (panel (d)) the green, black and blue lines correspond to $\left|A_{\rm{GS}\rightarrow 11}\right|$, $\left|\tilde{o}_{11}\right|$ and $\left|C_{11}^{(11)}\right|$ ($\left|A_{\rm{GS}\rightarrow 9}\right|$, $\left|\tilde{o}_{9}\right|$ and $\left|C_9^{(9)}\right|$), respectively. }
\label{F1}
\end{figure}

\begin{figure}
\centering
\includegraphics[width=0.49\textwidth]{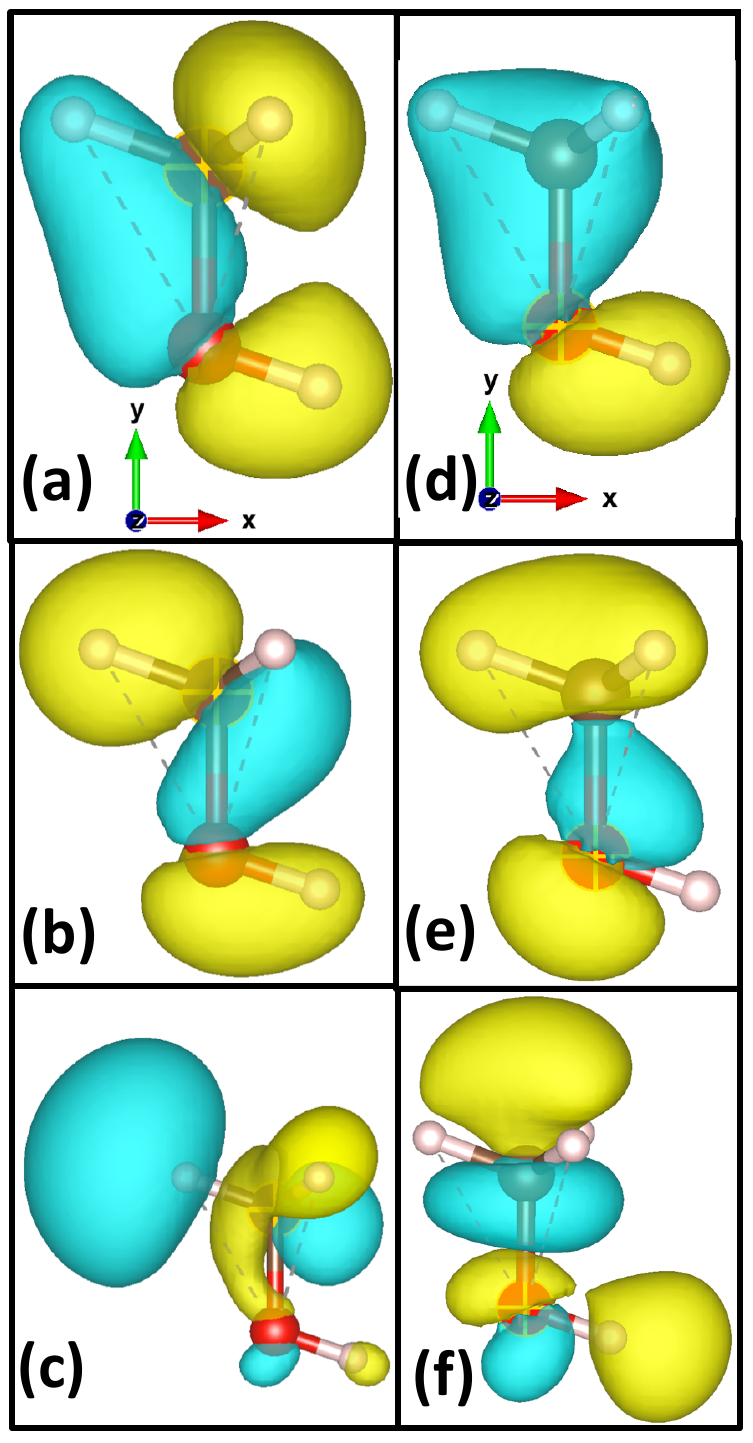}
\caption{Isovalue plots of occupied KS orbitals of the FCH-states of the methanol molecule resulting from core-ionization of specific atoms. The strongly contributing orbitals to the XAS intensity are indexed according to the energy ordering of their KS eigenvalues as (a) 3, (b) 5, (c) 11 (LUMO+2) for the C \textit{K} edge and (d) 2, (e) 5, (f) 9 (LUMO) for the O \textit{K} edge. The core-ionized atom is marked by a ``+'' sign.}
\label{MethanolOrbitalPlot}
\end{figure}

As stated concisely in Eq.~\ref{MBXAS_amplitude}, the MBXAS amplitude is essentially a sum of the terms
\begin{align}\label{C_l_definition}
    C_l^{(f)} = (-1)^{\gamma_l}\tilde{o}_l\braket{\Psi_{\rm{FCH}}^{+f-l+0}|\Psi_{\rm{GS}}} \textrm{ for } l=1, \hdots , N,f ,
\end{align}
for a given core-excited state $\ket{\Psi_{\rm{FCH}}^{+f}}$ labeled by core-excited (formerly unoccupied) FCH orbital index $f$.
We will now assess which $C_l^{(f)}$ are strong contributors to specific X-ray absorption transitions in methanol, by reference to Figs.~\ref{F1} and~\ref{MethanolOrbitalPlot}. 

At the C \textit{K}-edge in Fig.~\ref{F1}(a), the most intense XAS peak for photons polarized along the x-axis (see Fig.~\ref{MethanolOrbitalPlot} for the molecular orientation) occurs just below 290~eV and corresponds to transitions to $\ket{\Psi_{\rm{FCH}}^{+11}}$ for the core-excited C. As expected, the dominant contribution to this transition is $C_{11}^{(11)}$, defined by the $f=11$ orbital (FCH LUMO+2) that labels this excited state. However, in addition, we see in Fig.~\ref{F1}(c) that there are also significant transition amplitude contributions from $C_3^{(11)}$ and $C_5^{(11)}$ which involve occupied orbitals 3 and 5 from this FCH SCF. As expected, each of these orbitals (3, 5, 11) display noticeable $p_x$-character on the core-excited C atom, as shown in Fig.~\ref{MethanolOrbitalPlot}(a-c).

On the other hand, at the O \textit{K}-edge in Fig.~\ref{F1}(b), the most intense XAS peak for photons polarized along the y-axis occurs above 536~eV and corresponds to transitions to $\ket{\Psi_{\rm{FCH}}^{+9}}$ for the core-excited O. As before, the dominant contribution to this transition is $C_{9}^{(9)}$, defined by the $f=9$ orbital (FCH LUMO) that labels this excited state. However, in addition, we see in Fig.~\ref{F1}(d) that there are also significant transition amplitude contributions from $C_2^{(9)}$ and $C_5^{(9)}$ which involve occupied orbitals 2 and 5 from this FCH SCF. As expected, each of these orbitals (2, 5, 9) display noticeable $p_y$-character on the core-excited O atom, as shown in Fig.~\ref{MethanolOrbitalPlot}(d-f).

The computed absolute values of the valence-overlap between $\ket{\Psi_{\rm{FCH}}}$ and $\ket{\Psi_{\rm{GS}}}$ are 0.94 and 0.83 for the C and the O \textit{K}-edge of methanol, respectively. Since the excitation is expected to induce only negligible change in the tightly-bound core orbital, the aforementioned valence-overlap can be used as a good approximation for $\braket{\Psi_{\rm{FCH}}^{+0}|\Psi_{\rm{GS}}}$. As expected, owing to a higher overlap, the difference between $|C_{11}^{(11)}|$ and $|\tilde{o}_{11}|$ is smaller at the C \textit{K} edge than that between $|C_{9}^{(9)}|$ and $|\tilde{o}_{9}|$ at the O \textit{K} edge. However, the relative difference between the absolute values of the MBXAS and the single-particle amplitude, i.e., $\frac{|A_{\rm{GS}\rightarrow f}| - |\tilde{o}_f|}{|\tilde{o}_f|}$ (this is the relative difference between the green and black horizontal lines in Fig.~\ref{F1} (c) and (d)) is much higher for C (0.15) than it is for O (0.07), demonstrating that a lower overlap between the valence and the FCH occupied subspaces does not necessarily imply a larger difference between $|A_{\rm{GS} \rightarrow f}|$ and $|\tilde{o}_f|$. In general, the formally more accurate option is clearly to include all additional valence occupied contributions, although it is clear that the XAS is somewhat dominated by its single-particle contributions in this case. We will explore a counter-example below.

\subsection{Auxiliary Orbital}

\begin{figure}
\centering
\includegraphics[width=0.49\textwidth]{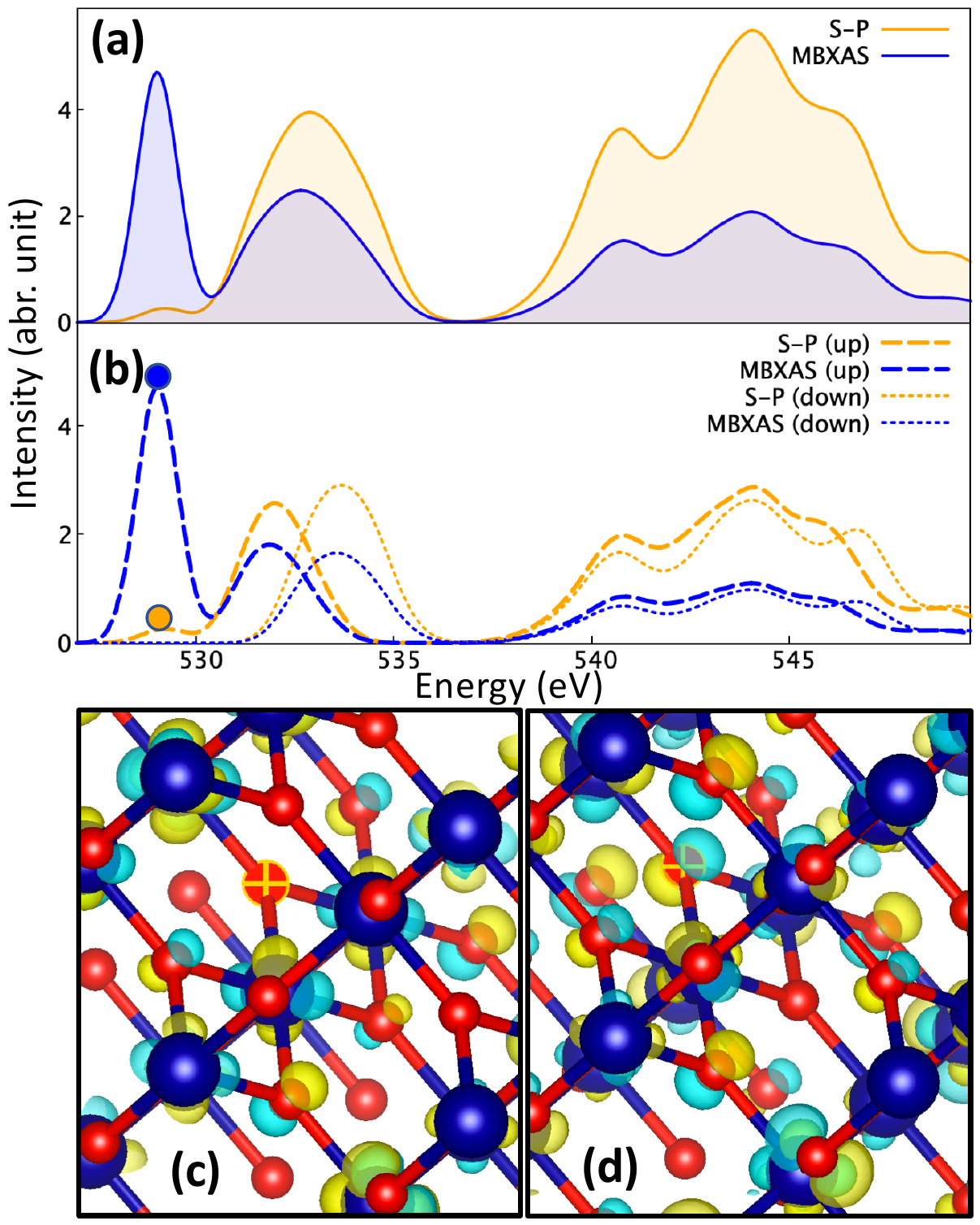}
\caption{Panel (a) shows the O \textit{K}-edge XAS of CrO$_2$ simulated using the single-particle (S-P) and MBXAS methods. Panel (b) shows separately the up and down spin contributions to spectra of both methods. The small orange (blue) circle corresponds to the first peak in the up-spin spectrum and is analysed using the isovalue plot of panel (c) (panel (d)) which shows, at the $\Gamma$ point, the orbital $\tilde{\phi}_f(\mathbf{r})$ ($\tilde{\phi}^{\rm{aux}}_f(\mathbf{r})$) that contributes most strongly to this peak. In this case, $f$ corresponds to the lowest unoccupied level of the FCH state. The core-ionized oxygen atom is marked by a ``+'' symbol in both lower panels.}
\label{CrO2SpecandIsov}
\end{figure}

It is important to note that the excitation described by MBXAS is an inherently many-body process with no physical single-particle analogue. However, the transition amplitude given by Eq.~\ref{MBXAS_amplitude} inspires the formulation of a hypothetical single-particle excitation of a core electron to an auxiliary orbital given by
\begin{align}\label{AuxiliaryExpansion}
\ket{\tilde{\phi}_{\rm{aux}}^f} =  \sum_{l=1,\hdots,N,f} (-1)^{\gamma_l} \braket{\Psi_{\rm{GS}}|\Psi_{\rm{FCH}}^{+f-l+0}}\ket{\tilde{\phi}_l}
\end{align}
such that the amplitude of this excitation equals that of the actual many-body one, i.e.,
\begin{align}\label{AuxiliaryAmplitude}
\tilde{o}^{\rm{aux}}_f = \braket{\tilde{\phi}_{\rm{aux}}^f| \hat{o}|\phi_0} = A_{\rm{GS}\rightarrow f},
\end{align}
as can be seen by comparison with Eq.~\ref{TransitionAmplitude}. 
The utility of the auxiliary orbital can be more easily appreciated by looking at the O \textit{K}-edge spectrum of CrO$_2$ which was discussed in detail in Ref~\cite{PhysRevB.97.205127}.

As shown in Fig.~\ref{CrO2SpecandIsov}(b), for the up-spin channel, the intensity of the O \textit{K}-edge first peak at 529~eV is severely underestimated by the single-particle approximation. This underestimation can be understood by examining, at the $\Gamma$ point, the corresponding excited single-particle orbital in the isovalue plots. The filled FCH orbital $f$, in this case, sits at the conduction band minimum (CBM), or first unoccupied orbital (analogous to LUMO at the $\Gamma$-point) of the FCH state. We see from Fig.~\ref{CrO2SpecandIsov}(c) that $\tilde{\phi}_{\rm{LUMO}}(\mathbf{r})$ has apparently no visible presence on the core-ionized O atom, which explains the almost vanishing intensity of this first peak in the single-particle spectrum. On the other hand, the associated auxiliary orbital $\tilde{\phi}^{\rm{aux}}_{\rm{LUMO}}(\mathbf{r})$ shown in Fig.~\ref{CrO2SpecandIsov}(d) exhibits substantial O-\textit{p} character on the core-ionized atom, due to inclusion of contributions from occupied valence orbitals in Eq.~\ref{AuxiliaryExpansion}, and hence gives rise to a drastically higher (and more accurate) MBXAS intensity.

\begin{figure}
\centering
\includegraphics[width=0.47\textwidth]{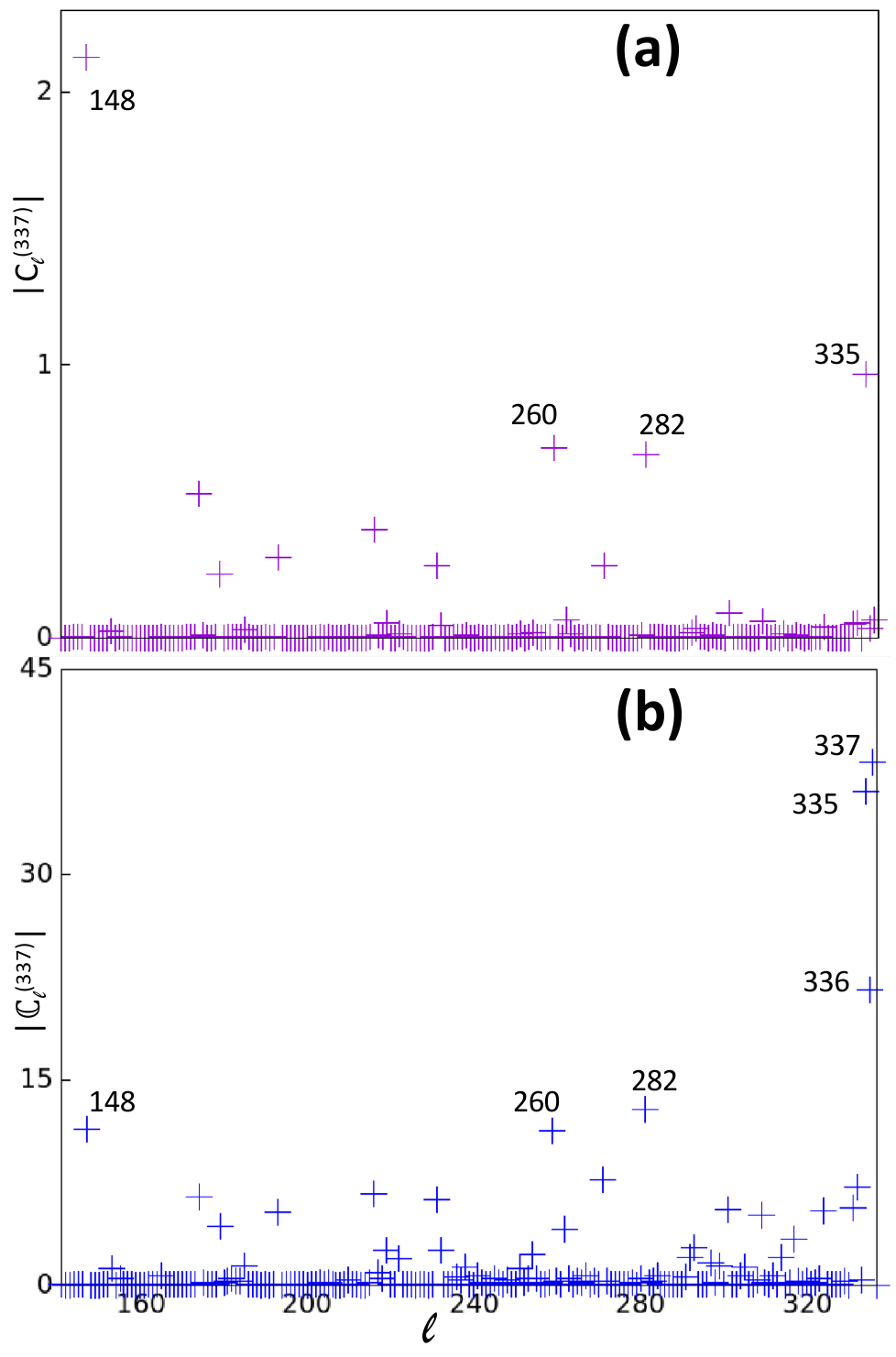}
\caption{For O \textit{K}-edge electronic transition at the $\Gamma$ point of CrO$_2$, panels (a) and (b) show, as a function of the FCH KS orbital indexed by $l$, the absolute values $\left|C_l^{(337)}\right|$ (see Eq.~\ref{C_l_definition})and $\left|\mathbb{C}_l^{(337)}\right|$, respectively, where the 337-th FCH orbital is the lowest unoccupied one.}
\label{C_l_and_mathbfC_l}
\end{figure}

\begin{figure}
\centering
\includegraphics[width=0.49\textwidth]{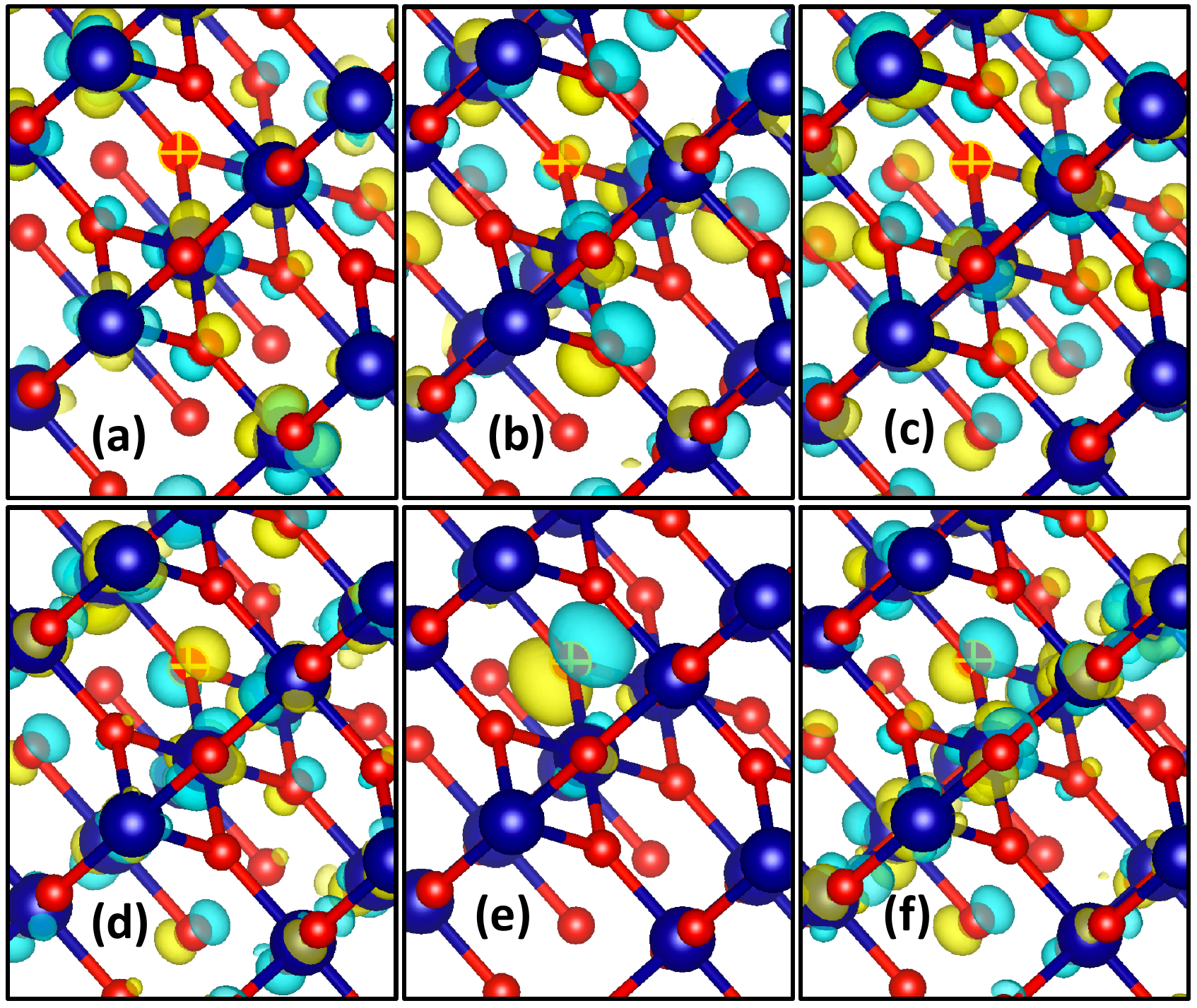}
\caption{Isovalue plots of $\Gamma$-point FCH KS orbitals contributing most strongly to the auxiliary orbital $\tilde{\phi}^{\rm{LUMO}}_f(\mathbf{r})$ shown in Fig.~\ref{CrO2SpecandIsov}(d). Panels (a) through (f) indicate decreasing contributions $\left|\mathbb{C}_l^{\rm{LUMO}}\right|$ starting from the largest:  (a) orbital 337 (LUMO), (b) 335 (HOMO-1), (c) 336 (HOMO), (d) 282, (e) 148 and (f) 260. The corresponding contributions can be found in Fig.~\ref{C_l_and_mathbfC_l}(b).}\label{Cro2AuxiliaryContributors}
\end{figure}

Fig.~\ref{C_l_and_mathbfC_l}(a) shows, as a function of the FCH orbital $\tilde{\phi}_l$, the absolute value of $C_l^{(\mathrm{LUMO})}$ (see Eq.~\ref{C_l_definition}), which is essentially the contribution to the first-peak MBXAS amplitude (Eq.~\ref{MBXAS_amplitude}) of the single-particle transition associated with the orbital $\tilde{\phi}_l$. On the other hand, Fig.~\ref{C_l_and_mathbfC_l}(b) shows the absolute value of $\mathbb{C}_l^{(\rm{LUMO})} = (-1)^{\gamma_l} \braket{\Psi_{\rm{GS}}|\Psi_{\rm{FCH}}^{+{\rm{LUMO}}-l+0}}$, which, as seen from Eq.~\ref{AuxiliaryExpansion}, serves as the contribution of $\tilde{\phi}_l$ to the auxiliary orbital $\tilde{\phi}^{\rm{aux}}_{\rm{LUMO}}$. As expected, due to the factor of $\tilde{o}_l$, these two contributions do not follow the same trend. Panel (a) of Fig.~\ref{C_l_and_mathbfC_l} clearly shows that the largest contribution to the MBXAS amplitude arises from the 148-th FCH orbital $\tilde{\phi}_{148}$, situated deep in the valence subspace. Predictably, $\tilde{\phi}_{148}$ exhibits strong $p$ character on the excited oxygen atom, as seen from panel~(e) of Fig.~\ref{Cro2AuxiliaryContributors}, which shows the isovalue plots of FCH orbitals with appreciable contribution in $\tilde{\phi}^{\rm{aux}}_{\rm{LUMO}}$. Conversely, as evident from  Fig.~\ref{C_l_and_mathbfC_l}(b), the largest contribution to the auxiliary orbital comes from the FCH LUMO, which, as already seen from Fig.~\ref{CrO2SpecandIsov}(c), has negligible $p$-character on the core-excited oxygen atom. For the auxiliary orbital $\tilde{\phi}_{\rm{LUMO}}^{\rm{aux}}$, the noticeable $p$ character on the excited atom can then be attributed to the cumulative contributions from the  O-\textit{p} character of various valence orbitals having appreciable $\mathbb{C}_l^{(\rm{LUMO})}$.

Note, from Eq.~\ref{GBA_AMPLITUDE} that, auxiliary orbitals can also be defined in the GB approach as
\begin{align}
    \ket{\tilde{\phi}_{\rm{aux}}^f} = \sum_{c>N} \braket{\Psi_{\rm{FCH}}^{+f}|\Psi_{\rm{GS}}^{+c-0}} \ket{\phi_c}.
\end{align}
However, due to the convergence issues discussed above, this might require the evaluation and manipulation of a large number of GS unoccupied orbitals, rendering the calculation formidably expensive.

\subsection{The $f^{(n)}$ contribution}

At this point, it is worth mentioning briefly the contribution of the so-called $f^{(n)}$ term~\cite{PhysRevB.97.205127} within the MBXAS formalism, whereby the final state
\begin{align*}
    \ket{\Psi_f^{(n)}} &{} = \ket{\Psi_{\rm{FCH}}^{+f_{n-1}+\hdots+f_1+f-h_{n-1}-\hdots-h_1}}\\
    &{}=(-1)^{\Lambda} \tilde{a}^\dagger_{f_{n-1}}\hdots \tilde{a}^\dagger_{f_1} \tilde{a}^\dagger_f \tilde{a}_{h_{n-1}} \hdots \tilde{a}_{h_1}\ket{\Psi_{\mathrm{FCH}}},\numberthis
\end{align*}
with $\Lambda=(nN-h_1-\hdots -h_{n-1})$, is obtained by generating, in addition to the core-excitation, $n-1$ number of electron-hole pairs outside the core-region, with the $m$-th conduction electron (valence hole) being indexed by $f_m$ ($h_m$). Without loss of generality, the orbital-indices are made to follow the order $f_{n-1} > \hdots >f_1>f>h_{n-1}>\hdots>h_1$. The core-excited state shown in Eq.~\ref{FinalState}, which contains no valence-hole, can now be recognized as $\ket{\Psi_f^{(1)}}$. 

Within the FB approach, the transition amplitude associated with $\ket{\Psi_f^{(n)}}$ can then be expressed as
\begin{align*}\label{fn_TransitionAmplitude}
&{}A_{\rm{GS}\rightarrow f}^{(n)}\\
    &{}=\braket{\Psi_f^{(n)}|\sum_l\tilde{o}_l\tilde{a}_l^\dagger \tilde{a}_0|\Psi_{\mathrm{GS}}}\\
    &{}=\sum_l (-1)^{\gamma_l^{(n)}} \tilde{o}_l \braket{\Psi_{\mathrm{FCH}}^{+f_{n-1}+\hdots+f_1+f-h_{n-1}-\hdots-h_1-l+0}|\Psi_{\mathrm{GS}}},\numberthis
\end{align*}
where
\begin{align*}\label{fn_TransitionAmplitude_2}
    \ket{\Psi_{\mathrm{FCH}}^{+f_{n-1}+\hdots+f_1+f-h_{n-1}-\hdots-h_1-l+0}}= (-1)^{\gamma_l^{(n)}}\tilde{a}_l\tilde{a}^\dagger_0\ket{\Psi_f^{(n)}},\numberthis
\end{align*}
with the term $\gamma_l^{(n)}$ depending on the relative position of $l$ with respect to $h_1 \hdots h_{n-1}$. Eq.~\ref{fn_TransitionAmplitude} and~\ref{fn_TransitionAmplitude_2} reveal that, within the FB approach, the GS unoccupied orbitals are not needed to compute the transition amplitude $A^{(n)}_{\rm{GS}\rightarrow f}$ for any value of $n$.  Similar to the special case of $f^{(1)}$ (see Eq.~\ref{AuxiliaryExpansion}), a general transition arising from the $f^{(n)}$ term can be associated with the auxiliary orbital
\begin{align}
    &{}\ket{\tilde{\phi}^{f^{(n)}}_{\rm{aux}}}\\ &{}= \sum_l (-1)^{\gamma_l^{(n)}} \braket{\Psi_{\rm{GS}}|\Psi_{\rm{FCH}}^{+f_{n-1}+\hdots+f_1+f-h_{n-1}-\hdots-h_1-l+0}}\ket{\tilde{\phi}_l}
\end{align}

As originally defined, we reiterate that within the MBXAS formalism, the contribution of $f^{(n)}$ for $n>1$ ignores interactions between the subsequent electron-hole pairs and, therefore, is really only accurate for studies of metallic systems. It is especially inaccurate for insulating/semiconducting systems which should exhibit significant exciton binding for additional valence electron-hole pairs.

\section{Conclusion}
In summary, in this article, we have presented an efficient representation of the determinant-based many-body formalism for simulating x-ray absorption spectroscopy, called MBXAS. We have shown that by recasting the relevant equations in terms of the FB orbitals, one can circumvent the need for calculating the unoccupied orbitals of the ground-state, leading to appreciable reduction in the computational expenses. Additionally, the reformulated expression for the transition amplitude does not suffer from convergence issues since it is now limited to a sum of a finite number of terms. Within the FB approach, the amplitude is expanded as a linear combination of independent-electron transitions from various core-excited state orbitals. Therefore, such a treatment facilitates a direct comparison of the MBXAS formalism with the so-called single-particle approximation. As we show analytically and with a representative example, the aforementioned comparison motivates the association of a given MBXAS transition to a purely single-particle electronic excitation from the core level to a hypothetical auxiliary orbital. A visual representation of this auxiliary orbital offers intuitive justification for the relevant absorption intensity.

\section{Acknowledgement}
This work was carried out at the Molecular Foundry (TMF),
Lawrence Berkeley National Laboratory (LBNL). Financial support for this work was provided by the Office of Science, Office of Basic Energy Sciences, of the U.S. Department
of Energy under Contract No. DE-AC02-05CH11231. Computational works were performed using the supercomputing resources of the National Energy Research Scientific Computing Center (NERSC) and the TMF clusters managed by the High Performance Computing Services Group, at LBNL.

\bibliographystyle{apsrev4-2}
%

\end{document}